\documentclass[aps,twocolumn,nofootinbib,superscriptaddress,preprintnumbers,pra,10pt]{revtex4-1}

\usepackage{float}
\usepackage{arydshln}
\usepackage{colortbl}
\usepackage{amsmath,amssymb}
\usepackage{dsfont} 
\usepackage{hyperref}
\usepackage{graphicx}
\usepackage{enumitem}
\usepackage{arydshln}
\usepackage{mathtools}
\usepackage{bbold}
\usepackage{soul}
\usepackage{slashed}
\usepackage{amsmath,bm}
\usepackage{yfonts}
\usepackage{lipsum}
\usepackage{slashed}
\usepackage{mathtools}
\usepackage{physics}
\usepackage{tikz}
\usepackage{tikz-feynman}

\makeatletter
\g@addto@macro\bfseries{\boldmath}
\makeatother

\newcommand{\be} {\begin{equation}}
\newcommand{\ee} {\end{equation}}
\newcommand{\bea} {\begin{eqnarray}}
\newcommand{\eea} {\end{eqnarray}}

\newcommand{\llpair}{\bar\ell\ell}

\renewcommand{\Re}{{\rm Re}}

\begin{document}

\preprint{ZU-TH-36/22}
\title{Explicit estimate of charm rescattering in $B^0 \to K^0 \llpair$}
 
\author{Gino Isidori }
\author{Zachary Polonsky}
\author{Arianna Tinari}

\affiliation{Physik-Institut, Universit\"at Z\"urich, CH-8057 Z\"urich, Switzerland}

\begin{abstract}
\vspace{5mm}
We analyze  $B^0 \to K^0 \llpair$
long-distance contributions induced by the 
rescattering of a pair of charmed and charmed-strange mesons. We present an explicit estimate of these contributions using an effective description in terms of 
hadronic degrees of freedom, supplemented by data on the $B^0 \to D^*D_s (D^*_sD)$ 
transition in order to reproduce the corresponding discontinuity in the 
$B^0 \to K^0 \llpair$
amplitude. The $D^* D_s (D^*_s D) K$ vertex is estimated using 
heavy-hadron chiral perturbation theory, obtaining an accurate description of the whole rescattering process in the low-recoil  (or high-$q^2$) limit.
We also present an extrapolation to the whole kinematical region introducing hadronic form factors. 
The explicit estimate of the leading $D^*D_s (D^*_sD)$ intermediate state leads to a long-distance amplitude which does not exceed a few percent
relative to the short-distance one. The consequences of this result
for the extraction of the short-distance coefficient $C_9$ are discussed.

\vspace{3mm}
\end{abstract}

\maketitle
\allowdisplaybreaks

\section{Introduction}\label{sec:intro}

Due to their strong suppression within the Standard Model (SM), exclusive and inclusive $b\to s\llpair$ decays are very interesting probes of short-distance physics. 
The exclusive $B\to~K^{(*)}\mu^+\mu^-$ modes have 
been measured with high accuracy in the last few years~\cite{LHCb:2013ghj,LHCb:2014cxe,LHCb:2015svh,CMS-PAS-BPH-22-005}. According to several analyses, data indicate a significant tension with the SM predictions~\cite{Alguero:2023jeh,Alguero:2018nvb,Gubernari:2020eft,Gubernari:2022hxn,Altmannshofer:2021qrr,Hurth:2020ehu,Wen:2023pfq,SinghChundawat:2022zdf, SinghChundawat:2022ldm}. The tension is particularly strong in the low-$q^2$ region, $q^2 = (p_\ell + p_{\bar{\ell}})^2$ being the invariant mass of the dilepton pair,
where the most stringent data-theory comparisons are currently made.
However, some tension is observed in the whole kinematical regime. Recent studies of the whole $q^2$ spectrum~\cite{LHCb:2023gpo,LHCb:2023gel,Bordone:2024hui}, taking into account the known singularities associated with the narrow charmonium resonances, indicate that data are well described by a sizable shift of the Wilson Coefficient $C_9$ of the semileptonic operator
\begin{equation}\label{eq:O9}
    \mathcal{O}_9 = (\bar{b}_L\gamma_\mu s_L)(\bar{\ell}\gamma^\mu \ell)\,,
\end{equation}
with respect to its SM value.\footnote{
Employing the standard notation 
for the $b\to s\bar\ell\ell$ effective Lagrangian~\cite{Blake:2016olu},
data favour $\Delta C_9\approx -1$  or $\Delta C_9/C_9^{\rm SM} \approx - 25\%$.} 
This shift is compatible with data-theory comparisons performed on the inclusive rates in the high-$q^2$ region~\cite{Isidori:2023unk,Huber:2024rbw}. However, the latter are affected by larger uncertainties and, at present, do not allow us to draw definite conclusions.

While there is no doubt that current data on the exclusive modes are well described by a modification of the value of $C_9$, it is more difficult to unambiguously identify the origin of this effect. A shift in the coefficient of the local operator would signal new short-distance dynamics, hence physics beyond the SM. However, as argued in \cite{Jager:2014rwa,Ciuchini:2022wbq,Ciuchini:2020gvn},  this apparent shift could be an effective description of unaccounted-for long-distance contributions of SM origin. In fact, an inaccurate estimate of the nonlocal matrix elements of the four-quark operators could simulate an effective change in $C_9$.

A precise prediction of short-distance dynamics entails a universal shift of $C_9$ 
--that is, a shift that is independent of both $q^2$ and the decay amplitude. The data are entirely consistent with this prediction, once the known singularities in the 
$q^2$ spectrum have been accounted for~\cite{Bordone:2024hui}.
 However, current uncertainties --both in the data and in local form factors-- prevent significant discrimination between the hypotheses of beyond-the-Standard Model (BSM) contributions and unaccounted-for long-distance contributions based solely on this aspect.

In this work, we aim to shed additional light on this issue by providing an estimate of long-distance effects associated with the rescattering of a pair of charmed and charmed-strange mesons, which have never been explicitly estimated so far. As pointed out in~\cite{Ciuchini:2022wbq}, these rescattering amplitudes are associated with physical thresholds 
which are not correctly reproduced in any of the available theory-driven estimates of the non-local matrix-elements of four-quark operators in $B\to~K^{(*)}\mu^+\mu^-$\footnote{Physical discontinuities associated to on-shell charm-quark states are present in the partonic estimate of the 
matrix-elements of four-quark operators in~\cite{Asatrian:2019kbk,Gubernari:2020eft}. However, the partonic calculation does not reproduce the correct physical thresholds and is subject to large duality violations in estimating the impact of the discontinuity. This is particularly true for the physical thresholds with the valence structure of a pair of charmed and charmed-strange mesons, which appear only at $O(\alpha_s)$ in the partonic calculation.
}.

To this purpose, we examine in detail the simplest decay mode, namely $B^0\to~K^0\llpair$, and the largest contributing individual two-body intermediate state to this mode, namely the one formed by a $D^*D_s$ or $D^*_sD$ pair. 
We estimate the rescattering amplitude using an effective description in terms of hadronic degrees of freedom
(i.e.~meson fields)
supplemented by data on the $B^0 \to D^*D_s (D^*_sD)$ decay. 
The $D^*D_s K$ vertex is estimated using 
heavy-hadron chiral perturbation theory (HHChPT), obtaining an accurate description of the whole rescattering process in the low-recoil (or high-$q^2$) limit. 
The extrapolation to the whole kinematical region is then performed introducing appropriate hadronic form factors.

The paper is organized as follows.
In Sec.~\ref{sec:model} we introduce the effective interactions used to perform the calculation in terms of mesonic fields.
In Sec.~\ref{sec:ffs} we discuss the modifications of the point-like vertices introduced in Sec.~\ref{sec:model} necessary to extrapolate the result to the whole kinematical region.
Analytical and numerical results for the $D^*D_s (D^*_sD)$ intermediate state are presented in Sec.~\ref{sec:res}. The consequences for the extraction of $C_9$, taking into account also additional intermediate states, are discussed in
Sec.~\ref{sec:disc}. The results are summarised in the Conclusions.

\section{Topologies and effective interactions}\label{sec:model}

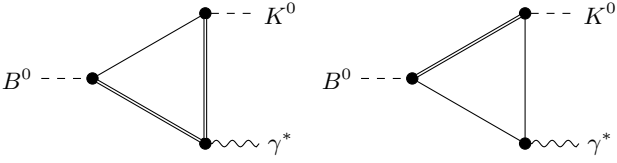
\begin{figure}[t]
        \centering
        \begin{tikzpicture}
                \begin{feynman}
                        \node (i1) {$B^0$};
                        \node[right=1cm of i1, dot] (i2);
                        \node[right=1.5cm of i2] (fake);
                        \node[above=0.866cm of fake, dot] (i3);
                        \node[below=0.866cm of fake, dot] (i4);
                        \node[right=1cm of i3] (j2) {$K^0$};
                        \node[right=1cm of i4] (j3) {$\gamma^*$};
                        \diagram*{
                                (i1) --[scalar] (i2) --[double] (i4) --[double] (i3) --[plain] (i2) ;
                                (i3) --[scalar] (j2) ;
                                (i4) --[photon] (j3) ;
                        };
                \end{feynman}
        \end{tikzpicture}
        \begin{tikzpicture}
                \begin{feynman}
                        \node (i1) {$B^0$};
                        \node[right=1cm of i1, dot] (i2);
                        \node[right=1.5cm of i2] (fake);
                        \node[above=0.866cm of fake, dot] (i3);
                        \node[below=0.866cm of fake, dot] (i4);
                        \node[right=1cm of i3] (j2) {$K^0$};
                        \node[right=1cm of i4] (j3) {$\gamma^*$};
                        \diagram*{
                                (i1) --[scalar] (i2) --[plain] (i4) --[plain] (i3) --[double] (i2) ;
                                (i3) --[scalar] (j2) ;
                                (i4) --[photon] (j3) ;
                        };
                \end{feynman}
        \end{tikzpicture}
        \caption{One-loop topologies considered in our analysis. Solid single lines denote charmed pseudoscalars ($D$ or $D_s$) and solid double lines denote charmed vectors ($D^*$ or $D^*_s$).}
        \label{fig:loops}
\end{figure}

Our goal is to  estimate the rescattering amplitudes 
associated with the two topologies in Fig.~\ref{fig:loops}, 
where an internal $D^*D_s (D^*_sD)$
pair can go on-shell.
With this aim in mind, we construct a model featuring the simplest effective interactions that are able to reproduce the discontinuities associated with these diagrams:
we describe the dynamics of $B^0$, $K^0$, $D$, $D^*$, $D_s$, and $D_s^*$ mesons using corresponding mesonic fields, denoted $\Phi_B$, $\Phi_K$, $\Phi_D$, $\Phi_{D^*}^\mu$, $\Phi_{D_s}$, and $\Phi_{D_s^*}^\mu$, respectively.

We stress that the model we consider is limited in scope to fulfill the above-stated goal, and is not meant to analyze rescattering amplitudes associated with different discontinuities (i.e.~different intermediate states). 
We focus specifically on these topologies as
they have been identified~\cite{Ciuchini:2022wbq} as the dominant contributions which are not captured by current theory-driven estimates of the non-local matrix elements of four-quark operators. Similar rescattering topologies have also been shown to produce sizable long-distance corrections in non-leptonic two-body decays~\cite{Cheng:2004ru}. On the other hand,
it is worth noting that multi-quark states, although relevant in hadron spectroscopy, should not play any role here (as additional intermediate states) since they are associated with subleading amplitudes in the $N_c$ large limit of QCD~(see e.g.~\cite{Allaman:2024vwn}).

Restricting the attention to the $D^*D_s (D^*_sD)$ intermediate state, we further neglect amplitudes induced by a dipole term of the type $D_{(s)} D^*_{(s)} \gamma$, which would generate additional topologies.
A discussion about the impact of the neglected topologies and additional discontinuities is presented 
in Sec.~\ref{sec:disc}.

To describe the dynamics of $D^{(*)}_{(s)}$  mesons close to their mass shell, we use the following effective Lagrangian,
which is determined only by the Lorentz transformation properties of the mesons and by gauge invariance under QED:
\begin{equation}
\begin{split}
        \mathcal{L}_{D,\text{free}} = & - \frac{1}{2}\big(\Phi_{D^*}^{\mu\nu}\big)^\dag\,\Phi_{D^*\,\mu\nu} - \frac{1}{2}\big(\Phi_{D^*_s}^{\mu\nu}\big)^\dag\,\Phi_{D^*_s\,\mu\nu} \\[0.5em]
        & +\big(D_\mu \Phi_D\big)^\dag\,D^\mu\Phi_D + \big(D_\mu \Phi_{D_s}\big)^\dag\,D^\mu\Phi_{D_s} \\[0.5em]
        &+m_D^2\big[\big(\Phi_{D^*}^\mu\big)^\dag\Phi_{D^*\,\mu} + \big(\Phi_{D^*_s}^\mu\big)^\dag\Phi_{D^*_s\,\mu}\big] \\[0.5em]
        & - m_D^2\big[\Phi_D^\dag\,\Phi_D + \Phi_{D_s}^\dag\Phi_{D_s}\big]+ \text{h.c.}\,.
\end{split}
\label{eq:QEDint}
\end{equation}
We have defined 
\begin{equation}
\begin{split}
    &\Phi_V^{\mu\nu} = D^\mu \Phi_V^\nu - D^\nu \Phi_V^\mu\,,\\[0.5em]
    &D_\mu \Phi = \partial_\mu \Phi + i\,e A_\mu \Phi\,,
\end{split}
\end{equation}
for $V = D^*, D_s^*$ and $e$ denotes the positron charge. Additionally, we have assumed $SU(3)$ light-flavor symmetry as well as heavy-quark spin symmetry to equate the masses of all the charmed-meson fields.

The weak $B\to D^*D_s (D^*_sD)$ transition  is described via the following effective Lagrangian
\begin{equation}
    \mathcal{L}_{BD} = g_{DD^*}\big(\Phi_{D_s^*}^{\mu\dag}\,\Phi_D\partial_\mu \Phi_B + \Phi_{D_s}^\dag \Phi_{D^*}^\mu \partial_\mu\Phi_B\big) + \text{h.c.}\,,
    \label{eq:LBD}
\end{equation}
where we again use heavy-quark spin symmetry to relate the coupling constants of the two terms.\footnote{The functional form of Eq.~(\ref{eq:LBD})  reproduces the functional dependence of the amplitude expected within 
na\"ive factorization starting from the non-leptonic weak effective Lagrangian. Given the kinematic constraints, the choice of this ansatz is irrelevant in the computation of the finite absorptive parts of the diagrams in Fig.~\ref{fig:loops}. } The value of the $g_{DD^*}$ coupling can be extracted from experimental data on $B$ decays. 
In order to facilitate a more straightforward comparison to the effective Lagrangian relevant to $b\to s \llpair$ decays, we redefine the coupling as
\begin{equation}
    g_{DD^*} = \sqrt{2}G_F\,|V_{tb}^*V_{ts}| m_B m_D\,\bar{g}\,,
    \label{eq:gDD_dec}
\end{equation}
where $V_{ij}$ denotes the elements  of the Cabibbo-Kobayashi-Maskawa (CKM) matrix. Beside the obvious dependence from $G_F$ and the $V_{ij}$ elements, the dependence on 
$m_B$ and $m_D$ in Eq.~(\ref{eq:gDD_dec}) is such that 
$\bar{g}$ is dimensionless and the $B\to D^*D_s$ rate computed using $\mathcal{L}_{BD}$ is not singular in the limit $m_D \to 0$. 
Using the average of $B\to D^*D_s$ and $B\to D^*_s D$ branching fractions from Ref.~\cite{PDG}, and
$B^0$ and $D^0$ masses in  (\ref{eq:gDD_dec}),
we find
\begin{equation}
    \bar{g}\approx 0.04\,.
\end{equation}
In general, the $g_{DD^*}$ coupling can have a complex phase that we cannot determine from available data. For simplicity, in the explicit calculations 
 we assume this coupling to be real, but we will come back to discussing the impact of this phase in Sec.~\ref{sec:disc}.

The remaining vertices necessary to estimate the rescattering process are those related to the kaon emission from the charmed mesons. We estimate these heavy-heavy-light vertices using the heavy-hadron chiral perturbation theory Lagrangian, obtaining 
\begin{equation}\label{eq:DKVerts}
    \mathcal{L}_{DK} = \frac{2ig_\pi m_D}{f_K}\big(\Phi_{D^*}^{\mu\dag}\Phi_{D_s}\partial_\mu\Phi_K^\dag - \Phi_D^\dag\Phi_{D_s^*}^\mu\partial_\mu\Phi_K^\dag\big) + \text{h.c.}\,,
\end{equation}
where $f_K = 155.7(3)$ MeV~\cite{PDG} is the kaon decay constant and $g_\pi \approx 0.5$~\cite{Becirevic:2012pf}.
By construction, the HHChPT approximation is only valid for soft kaon emission. In the processes we are interested in, this occurs for $q^2$
near the kinematic endpoint ($q^2_{\text{max}} \approx m_B^2$). This is the region where our estimate of the rescattering amplitude is more reliable. However, we also present an extrapolation to lower $q^2$ values by means of an appropriate form factor to account for the hard recoil momenta at the $KDD^*$ vertex (see Sec.~\ref{sec:ffs}).

For simplicity, we report analytic results in the $SU(3)$-symmetric limit
(i.e.~setting $m_K = 0$). From an explicit numerical calculation of the amplitude with the physical kaon mass, we find that associated $SU(3)$-breaking effects amount to a 
correction varying between $10\%$ and $20\%$.

\section{Form Factors}\label{sec:ffs}

\begin{figure}[t]
    \centering
    \includegraphics[width = 0.85\linewidth]{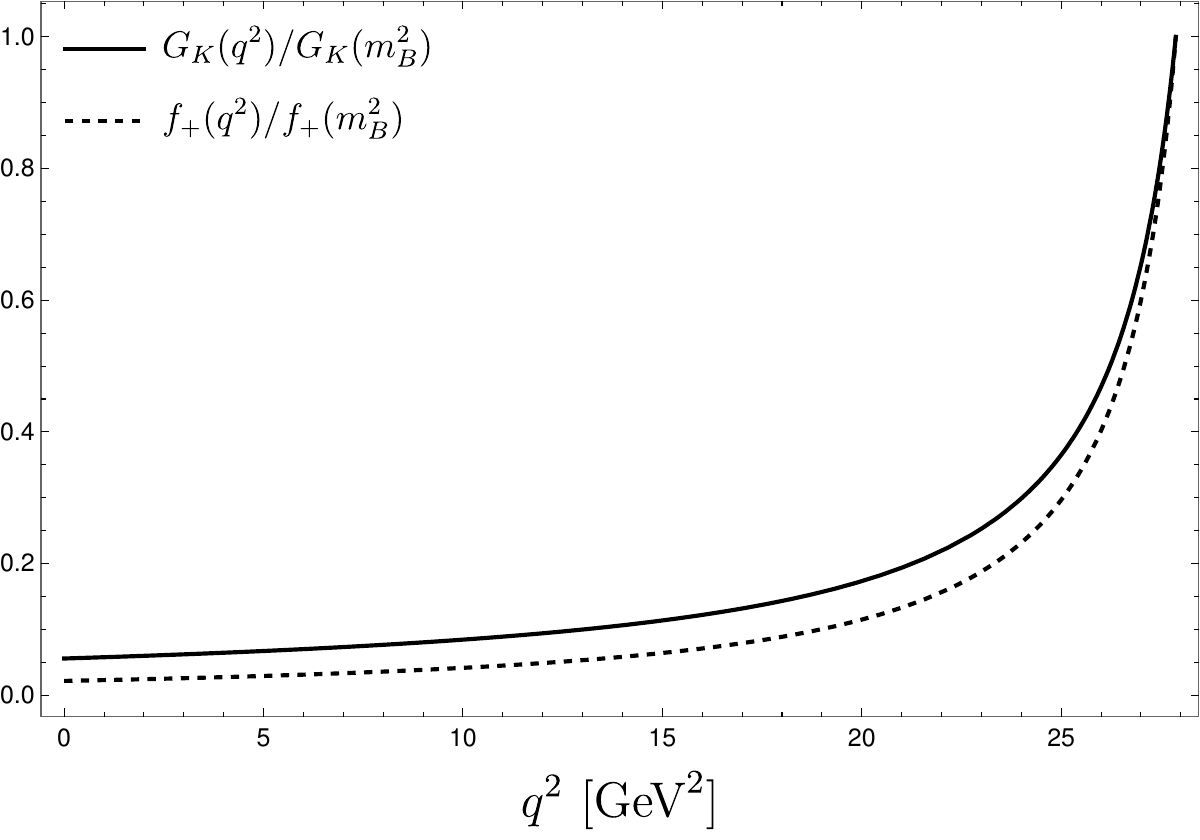}
    \caption{Scaling behavior of the form factor introduced for the $KDD^*$ interaction, Eq.~\eqref{eq:kaonFF}, compared to that of $f_+(q^2)$ from the lattice results in Ref.~\cite{Parrott:2022rgu}. Both form factors are normalized to their endpoint values.}
    \label{fig:FFs}
\end{figure}

In order to obtain a reliable estimate of the rescattering amplitude over the entire kinematical range, we must take into account the fact that the hadrons are not well-described by fundamental fields far from their mass shell. 
Given the kinematics of the process,  at high $q^2$ we need to introduce an appropriate electromagnetic form factor to correct the point-like QED vertices derived from  $\mathcal{L}_{D,\text{free}}$.
At low-$q^2$, the description of the 
$KDD^*$ vertex needs to be modified. 

As far as the electromagnetic form factor is concerned, we modify the point-like vertices in (\ref{eq:QEDint}) as follows
\begin{equation}
e \to  e F_V(q^2)\,, 
\quad F_V(q^2)=\left\{ \begin{array}{ll}  1\,, \qquad & q^2=0\,, \\
 \sim q^{-2}\,, \qquad & q^2 \gg m_D^2\,.
 \end{array} 
 \right.
\end{equation}
The normalization of $F_V(q^2)$ at $q^2=0$ 
is a consequence of the 
conservation of the vector current (i.e.~the conservation of the electric charge), while 
the asymptotic behavior  for $q^2 \gg m^2_D$ 
follows from perturbative QCD~\cite{Brodsky:1989pv}.
Both these conditions are fulfilled by the 
vector meson dominance (VMD) ansatz 
\begin{equation}
\label{eq:emFF}
F_V(q^2) = \frac{m_{J/\psi}^2}{m_{J/\psi}^2 - q^2}\,,
\end{equation}
that we employ in our analysis.
This ansatz is known to work well, phenomenologically,  
and it can be justified in the large $N_c$ limit of QCD~\cite{Peris:1998nj}. In such limit, the asymptotic behavior of the form factor is obtained summing over an infinite set of narrow vector resonances, leading to the following decomposition  (see e.g.~\cite{Dominguez:2001zu})
\begin{equation}
    F_V(q^2) = \sum_i c_i \frac{ m^2_i}{ m^2_i -q^2}\,,
     \label{eq:FV-param}
\end{equation}
where the condition $F_V(0)=1$ implies $\sum_i c_i=1$.
In our case, the tower of resonances in Eq.~(\ref{eq:FV-param}) is  dominated by the narrow charmonium states (as 
implied by perturbative QCD in the high $q^2$ region~\cite{Brodsky:1989pv} and confirmed by the structure of dilepton peaks in $B\to K\ell^+\ell^-$).\footnote{Light-quark resonances play a non-negligible role at low $q^2$; however, we are not interested in a precise local description of the form factor as a function of $q^2$, rather in its behavior smeared over wide $q^2$ intervals. In the low $q^2$ region this is constrained by the condition
$F_V(0)=1$.}
Considering only such states, 
neglecting their mass splitting (which is a good approximation both at low and high $q^2$), and imposing $\sum_i c_i=1$, leads to the expression in Eq.~(\ref{eq:emFF}).
Actually for $q^2 \gg m^2_{J/\Psi}$ the result in (\ref{eq:emFF}) slightly overestimates the result obtained in perturbation theory~\cite{Brodsky:1989pv}, 
providing a safe approximation in view of an 
estimate of the maximal size of the amplitude.

From the above discussion we can thus conclude that the change of sign of $F_V(q^2)$ in the low- and high-$q^2$ regions 
(i.e.~for  $q^2 \ll m^2_{J/\Psi}$ and 
$q^2 \gg m^2_{J/\Psi}$) is a 
general feature dictated by
properties of QCD. As we will discuss below, this has important phenomenological consequences for the rescattering effects.

\medskip

The modification of the $KDD^*$ vertex is less straightforward. Actually what we need to introduce is not a  form factor for the $KDD^*$ vertex (which is not well defined given at least one of the hadrons will be far off-shell), but rather a $q^2$-dependent correction of the 
whole $DD^* \to \bar\ell\ell K$ amplitude.
The first point to note is that that the Lagrangian in Eq.~\eqref{eq:DKVerts} leads to a $K$-emission amplitude that grows as $E_K/f_K$ with the kaon energy ($E_K$). This behavior is correct in the soft-kaon limit (Goldstone-boson emission) but needs to be corrected for
$E_K > f_K$, otherwise it would violate unitarity. A similar conclusion is reached by noting that the apparent $1/f_K \sim 1/\Lambda_{\rm QCD}$ behavior of the amplitude can appear only in the region of kaon momenta of $O(\Lambda_{\rm QCD})$. To address both of these issues, we use a form factor that makes the replacement
\begin{align}
\label{eq:kaonFF}
    \frac{1}{f_K} &\to \frac{1}{f_K} G_K(q^2)\,,
    \\
    G_K(q^2) &= \frac{1}{1 + E_K(q^2)/f_K}
    = \frac{2 m_B f_K}{2m_B f_K + m_B^2 - q^2}\,.
    \nonumber
\end{align}
By construction, $G(m_B^2)=1$, hence no correction is applied at the kinematical endpoint ($q^2_{\text{max}}=m_B^2$
in the $m_K\to0$ limit) when the kaon is soft
and we can trust the HHChPT result.
On the other hand, the 
denominator of $G_K(q^2)$ quickly grows further from the endpoint, compensating for the growth of the amplitude with the kaon energy. 
With this choice, at large kaon 
energies the emission amplitude approaches 
a constant. The ansatz in 
Eq.~(\ref{eq:kaonFF})
cannot be justified directly from QCD; 
however, it is interesting to note that the postulated functional form  of $G_K(q^2)$, normalized to its endpoint value, is in good agreement with that of $f_+(q^2)$, i.e.~the 
$B^0\to K^0$ vector factor, which is determined from lattice  QCD~\cite{Parrott:2022rgu} 
(see Fig.~\ref{fig:FFs} ). 
This is a useful consistency check since $f_+(q^2)$ 
shares similar features with $G_K(q^2)$, namely a
maximum in the limit of soft-kaon emission 
(kinematical endpoint) and 
an $O(\Lambda_{\rm QCD}/E_K)$-suppression far from this limit. 
As we will discuss in more detail in Sec.~\ref{sec:disc}, a similar scaling between 
$f_+(q^2)$ and $G_K(q^2)$ results in a roughly 
$q^2$-independent, $C_9$-like contribution, in both the high- and low-$q^2$ regions. We thus achieve the scenario envisaged in 
 Ref.~\cite{Ciuchini:2022wbq, Ciuchini:2020gvn} for this type of 
 rescattering contributions.

\section{Results}\label{sec:res}

\begin{table}[t]
    \centering
    \begin{tabular}{|c|c|}
    \hline
        parameter & value  \\
         \hline
         $m_B$ & 5.27966(12) GeV \\
         $m_D$ & 1.96506(11) GeV\\
         $m_{J/\psi}$ & 3.096900(6) MeV \\
         $g_\pi$ &  0.5\\
         $f_K$ & 155.7(3) MeV\\ 
         $|V_{ts}|$ & 0.041(1)  \\
         $G_F$ & 1.1663788(6)$\cdot 10^{-5}$ GeV$^{-2}$\\
         $C_9(m_b)$ & 4.114(14)\\
         \hline
    \end{tabular}
    \caption{Numerical inputs.}
    \label{tab:inputs}
\end{table}

\begin{figure}[t]
    \centering
    \includegraphics[width=0.9\linewidth]{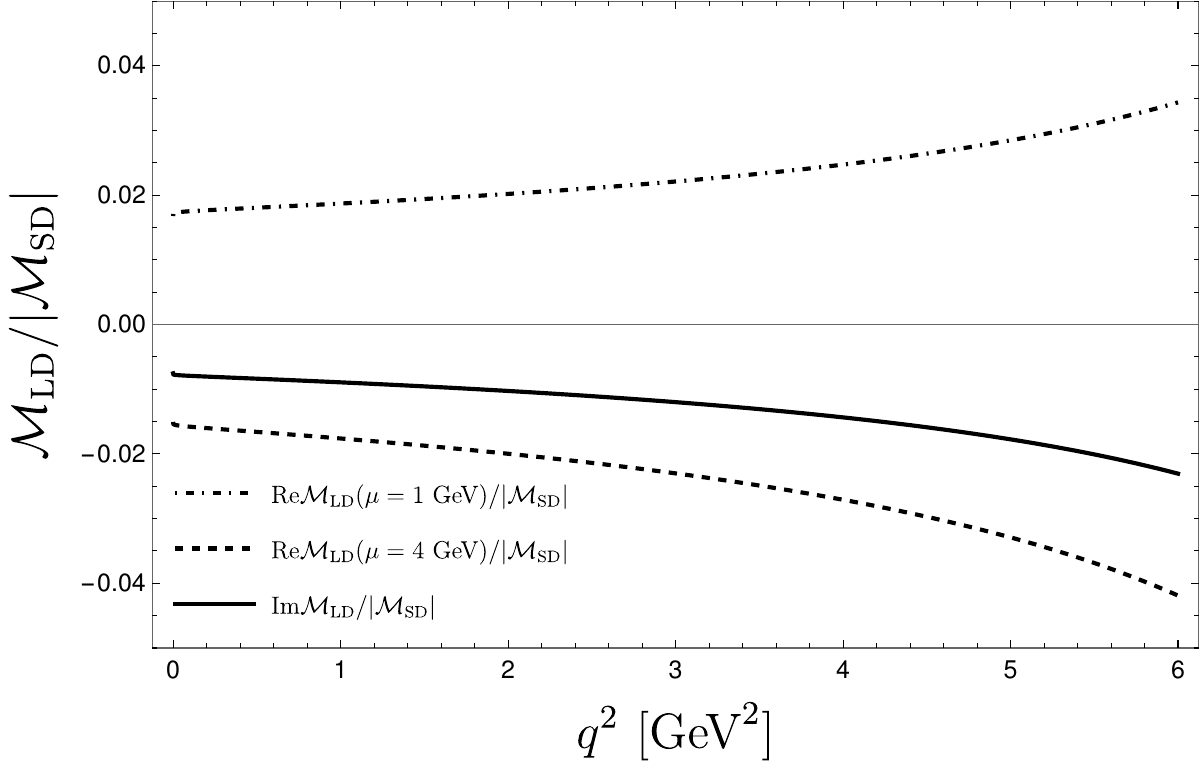}\\
    \vspace{0.5cm}
    \includegraphics[width=0.9\linewidth]{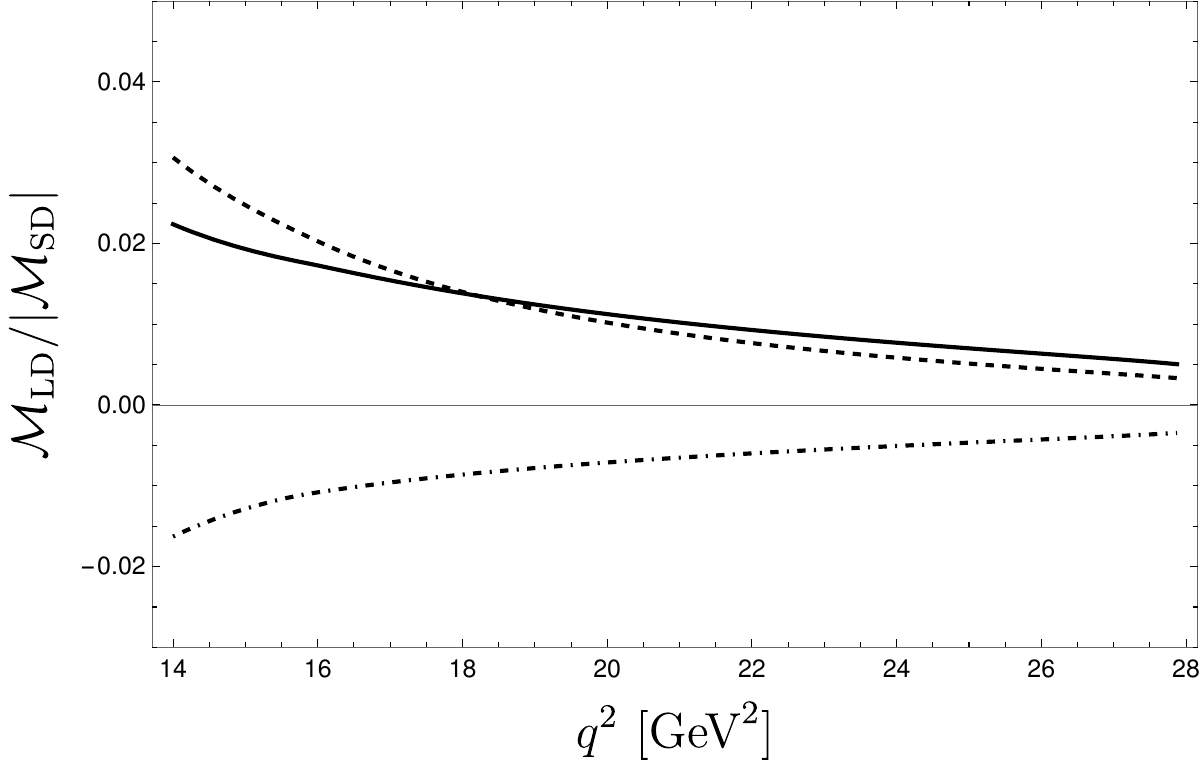}
    \caption{Ratio between charm rescattering contributions calculated in terms of hadronic degrees of freedom,
    to the matrix element without considering rescattering effects in the low-$q^2$ (top) and the high-$q^2$ (bottom) regions.}
    \label{fig:matel}
\end{figure}

We now have all the ingredients to estimate 
charm rescattering effects in $B^0\to K^0\llpair$
associated to the two topologies in Fig.~\ref{fig:loops}.
As anticipated, we estimate the corresponding one-loop diagrams using the effective Lagrangians introduced in Sec.~\ref{sec:model}. Their structure implies that the two topologies in Fig.~\ref{fig:loops} constitute 
an independent set of gauge-invariant diagrams. In the $SU(3)$-symmetric limit, diagrams obtained by replacing $D\leftrightarrow D_s$ and $D_s^*\leftrightarrow D^*$ are identical, thereby resulting in an overall factor of two for each topology.

As expected, the sum of diagrams features an ultraviolet divergence which we discard, employing an $\overline{\text{MS}}$-like renormalization scheme. This does introduce a scale-(and renormalization-scheme) dependence in our final result, that we can turn into a tool to estimate the associated uncertainty. In principle, additional finite counterterms can be included which alter the resulting piece of the matrix element that we calculate. However, in a full calculation, unphysical finite counterterms must exactly cancel with those arising from the short-distance part of the matrix element. Since we are only interested in an estimation of long-distance effects, we neglect such short-distance contributions.

In the $SU(3)$- and heavy-quark spin-symmetric limits, the contribution to the $B^0\to K^0\llpair$ matrix element arising from the loops in Fig.~\ref{fig:loops} is 
\begin{equation}\label{eq:ourRes}
    \begin{split}
        \mathcal{M}_{\text{LD}} &= -\frac{e g_{DD^*} g_\pi  F_V(q^2) G_K(q^2) }{8\pi^2 f_K m_D } (p_B\cdot j_{\text{em}})
        \\[0.5em]
        &\times\Big[\big(2 + L_\mu\big)
        - \delta L(q^2, m_B^2, m_D^2)\Big],
    \end{split}
\end{equation}
where $j_{\text{em}}^\mu = -e\,\bar{\ell}\gamma^\mu \ell$ is the conserved electromagnetic current of the dilepton pair. 
We defined the renormalization scale-dependent logarithm 
\be
L_\mu~=~\log(\mu^2/m_D^2)\,, 
\ee
as well as
\begin{equation}
    \begin{split}
        &\delta L(q^2, m_B^2, m_D^2) = \frac{L(m_B^2, m_D^2) - L(q^2, m_D^2)}{q^2 - m_B^2}\,, \\[0.5em]
        &L(x, y) = \log\Bigg(\frac{2 y - x + \sqrt{x(x - 4y)}}{2y}\Bigg) \\[0.5em]
        &\times \Bigg[\sqrt{x(x - 4 y)} + y \log\Bigg(\frac{2 y - x + \sqrt{x(x - 4y)}}{2y}\Bigg)\Bigg]\,.
    \end{split}
\end{equation}
It is worth stressing that while individual contributions to the amplitude 
exhibit poles at $q^2= 0$, associated with the photon propagator, the final result is regular at $q^2= 0$, as expected by gauge invariance.  

To determine the size of this long-distance contribution, we compare $\mathcal{M}_{\text{LD}}$ with the 
 short-distance amplitude generated by the operator $\mathcal{O}_9$ 
 in Eq.~(\ref{eq:O9}), which has exactly the same Lorentz structure. 
Using the normalization of $b\to s \bar\ell\ell$ effective Lagrangian in~\cite{Blake:2016olu} the short-distance contribution reads
\begin{equation}
\label{eq:noLDCharm}
    \mathcal{M}_{\rm SD} = \frac{4 G_F}{\sqrt{2}}\frac{e}{16\pi^2}V^*_{tb} V_{ts} (p_B\cdot j_{\text{em}})
    f_+(q^2)
    (2 C_9 ), 
\end{equation}
where $f_+(q^2)$ is the $B\to K$ 
vector form factor~\cite{Parrott:2022rgu}.

Numerical comparisons of Eqs.~\eqref{eq:ourRes} and \eqref{eq:noLDCharm} are shown in Fig.~\ref{fig:matel}, where the dispersive and absorptive parts of Eq.~\eqref{eq:ourRes} are plotted separately in both the high- and low-$q^2$ regions. We additionally show the dispersive part of the matrix element at two different values of the renormalization scale, $\mu = 1$ GeV and $\mu = 4$ GeV. The numerical values of the inputs used in the calculation are shown in Table \ref{tab:inputs}. 

We stress that the absorptive part of the amplitude 
is independent of the renormalization scheme used and, at least in the high-$q^2$ region, can be considered as a model-independent result. This part of the amplitude is determined by the analytic discontinuity occurring in the kinematical region where the internal mesons go on-shell. At high $q^2$, the coefficient of the discontinuity depends on on-shell amplitudes determined either from data and/or from reliable theoretical hypotheses. 
As an independent check, we have calculated separately 
these discontinuities, finding perfect agreement with the results in Eq.~\eqref{eq:ourRes} which follows from the explicit loop calculation.

In principle, rather than computing the loop amplitude, the dispersive part of the amplitude could have been determined by a dispersion relation starting from the (model-independent) result for the discontinuity. This method was recently applied to the case of intermediate light-quark states~\cite{Mutke:2024tww}.  Proceeding that way, the ambiguity of the dispersive part, which in the loop calculation manifests itself via the ultraviolet divergence, would be hidden in the subtraction constant of the dispersion relation
(and in the extrapolation of the absorptive part in kinematical regions where we have no data to constrain it).
The fact that, varying the renormalization scale, we find a dispersive part very similar in size to the absorptive one is in good agreement with what is found e.g.~in $s$-channel $\pi\pi$ scattering (see e.g.~\cite{Colangelo:2000zw}), where the subtraction constant is known and the dispersion relation can be solved exactly. Similar dispersive and absorptive parts are also found in the contributions due to intermediate charmed meson states in $B \to K \pi $ decays in \cite{Isola:2001ar}.
We cannot exclude the case that in the process we consider the unknown subtraction constant is unexpectedly large; however, in all known cases a large subtraction constant is associated with a nearby strongly coupled resonance,  such as the $\rho$ in $p$-channel $\pi\pi$ scattering (as expected in the large $N_c$ limit). Since we have no indication that this happens in the present case, we stick to the natural indication provided by the explicit loop calculation.

\section{Discussion}\label{sec:disc}

As for any long-distance contribution to $B\to K\bar\ell\ell$ at $O(\alpha_{\rm em})$, 
we can encode the effect of  $\mathcal{M}_{\text{LD}}$
in Eq.~(\ref{eq:ourRes}) via a $q^2$--dependent shift of $C_9$. Doing so 
leads to 
\be 
\label{eq:dC9}
    \delta C^{\rm LD}_{9,DD^*}(q^2,\mu) = 
    \bar{g}\, \Delta(q^2)\Big[2 + L_\mu - \delta L(q^2, m_B^2, m_D^2)\Big]\,,
\end{equation}
where
\begin{equation}
    \Delta(q^2) =  -\frac{  g_\pi  m_B F_V(q^2) G_K(q^2) }{ 2 f_K  f_+(q^2) }\,.
\end{equation}
The quantity $\Delta(q^2)$ is the combination of hadronic parameters controlling the rescattering process.
As already discussed, this is estimated more reliably at $q^2_{\text{max}}=m_B^2$, where $\Delta(m^2_B)\approx 1.5$.
From this normalization, we deduce that the natural size of this rescattering amplitude, relative to the short-distance one, is
\be
\left| \frac{\delta C^{\rm LD}_{9,DD^*}}{C_9} \right| = O(1) \times 
\left| \frac{\bar g}{C_9} \right| =
O(1\%)\,.
\ee 

The choice of the hadronic form factors makes this correction relatively flat in $q^2$, far from the 
narrow charmonium region. 
Averaging Eq.~\eqref{eq:dC9} over the conventional low- and high-$q^2$ regions, $q^2\in [0,6]$ GeV$^2$ and $q^2\in [14$ GeV$^2, m_B^2]$, respectively, leads to
\begin{equation}\label{eq:dC9avg}
    \begin{split}
        \delta\bar{C}_{9,DD^*}^{\text{LD,low}}(\mu) &= - 0.003 - 0.059\,i - 0.156\log\Big(\frac{\mu}{m_D}\Big)\,,\\[0.5em]
        \delta\bar{C}_{9,DD^*}^{\text{LD,high}}(\mu) &= 0.009 + 0.053\,i + 0.063\log\Big(\frac{\mu}{m_D}\Big)\,.
    \end{split}
\end{equation}
Since $\Re\,\delta L(q^2, m_B^2, m_D^2) \approx 2$, the real parts in Eq.~\eqref{eq:dC9avg} turn out to be  particularly suppressed for $\mu=m_D$. This accidental suppression is lifted varying $\mu$ in the range $[1,4]$~GeV, where the real and imaginary parts become of the same order. Doing so leads to
\be
\label{eq:dC9tot}
   | \delta \bar{C}^{\rm LD}_{9,DD^*} | \leq 0.11\,.
\end{equation}
The relative phase difference between short- and long-distance contributions depends on the unknown phase of the $g_{DD^*}$ coupling (so far we have set it to be real for simplicity). Assuming a maximal interference, the result in Eq.~(\ref{eq:dC9tot}) implies a maximal correction to $C_9$ in the low- or high-$q^2$ region of $2.5\%$.

An interesting point to note is that while 
$\delta C^{\rm LD}_{9,DD^*}$ has a mild $q^2$-dependence in the low- and high-$q^2$ regions (hence it can effectively mimic a short-distance contribution in each region), the sign is opposite in the two cases (regardless of the phase of $g_{DD^*}$).
This is an unavoidable consequence of the structure of the electromagnetic form factor in Eq.~(\ref{eq:emFF}) and is mildly affected by other hypotheses. Hence comparing the extraction of $C_9$ at low- and high-$q^2$, as advocated in~\cite{Bordone:2024hui}, provides a useful data-driven check for such long-distance contributions. To this purpose, we recall that present data do not indicate a statistically significant difference~\cite{Bordone:2024hui}.

\begin{table}[t]
    \centering
    \begin{tabular}{|c|c|}
    \hline
        $B^0$ Decay & $\mathcal{B}(B^0\to X)\times 10^3$  \\
         \hline
         $D^{*}D_s$ & $8.0\pm 1.1$\\
         $D D_s^*$ & $7.4\pm 1.6$\\
         $D^*D_s^*$ & $17.7\pm1.4$\\
         $D D_{s0}(2317)$ &  $1.06\pm1.6$\\
         $D^* D_{s1}(2460)$ &  $9.3\pm2.2$\\
        $D^* D_{s1}(2536)$ & $0.50 \pm 0.14$ \\
         $D D_{s2}(2573)$ &  $(3.4\pm1.8)\times 10^{-2}$\\
         $D^* D_{s2}(2573)$ &  $<0.2$\\
         $D D_{s1}(2700)$ &  $0.71\pm0.12$\\
         \hline
    \end{tabular}
    \caption{List of additional charm-strange $B^0$ decay modes which allow 
    parity-conserving strong interactions with the $K^0$. All values come from Ref.~\cite{PDG}, and the $D_{sJ}(2460)$ is assumed to be $J^P=1^+$ as indicated in Ref.~\cite{Belle:2003guh}.}
    \label{tab:BDDdecays}
\end{table}

So far we focused only on the $D^*D_s (D^*_sD)$ intermediate state. 
In principle, there exist additional intermediate states with $\bar{c}c\bar{s}d$ valence structure that can lead to a similar rescattering amplitude. The relevant two-body 
decay modes  which allow for parity-conserving strong interactions with the $K^0$ are reported in Tab.~\ref{tab:BDDdecays}.\footnote{
 We do not consider baryonic modes since the leading channel, $\mathcal{B}(B^0\to \overline{\Xi}_c^-\Lambda_c^+) = (1.2\pm0.8)\times 10^{-3}$~\cite{PDG}, is $\lesssim 10\%$ compared to the sum of $B\to D^* D_s$ and $B\to DD^*_s$ modes.}
To provide a rough   estimate of the impact of these additional states, we normalize the $B^0\to X_{\bar{c}c\bar{s}d}$ rates to the $B^0\to D^*D_s+
DD^*_s$ one, assuming that each gauge-invariant subset of diagrams roughly scales with the size of the corresponding $B^0\to X_{\bar{c}c\bar{s}d}$ amplitude with respect to those we have calculated.
Doing so, using the values shown in Tab.~\ref{tab:BDDdecays}, we 
derive the following estimate of the 
maximal multiplicity factor
\begin{equation}\label{eq:multiplicity}
    \begin{split}
    \mathcal{N} &= \frac{\sum_X\mathcal{M}(B^0\to X)}{\mathcal{M}(B^0\to D^*D_s) + \mathcal{M}(B^0\to DD^*_s)} \\[0.5em]
    &\approx \frac{1}{2}\sum_X\sqrt{\frac{\mathcal{B}(B^0\to X)}{\mathcal{B}(B^0\to DD_s^*)}}\approx 3\,,
    \end{split}
\end{equation}
associated with the additional modes. While this estimate is admittedly rough, it is also reasonably conservative, being based on the assumption that all possible contributing intermediate states add coherently in the final result. Using the 
multiplicity factor in (\ref{eq:multiplicity}),
we estimate the maximal correction to $C_9$ as
\begin{equation}\label{eq:findC9}
    |  \delta C_{9}^{\rm LD} | 
    \leq
    \mathcal{N} 
    | \delta \bar{C}^{\rm LD}_{9,DD^*} |  \leq 0.33\,.
\end{equation}
 Eq.~\eqref{eq:findC9} neglects several factors such as $SU(3)$-breaking effects ($\lesssim 30 \%$), higher-mass charmonium resonances ($N_c$-suppressed), baryonic modes ($\lesssim 10\%$), and higher-multipole photon couplings ($m_c$-suppressed). Although each of these effects are expected to be small individually, they can possibly lead to a larger-than-expected enhancement, relaxing the bound in Eq.~\eqref{eq:findC9}. A further comment is warranted on the $D^*D_{s1}(2460)$ mode, as this mode can couple to the kaon in $S$-wave as well as receiving an enhancement from the branching ratio of $B\to D^*D_{s1}(2460)$. The latter is accounted for in our multiplicity factor, but since the $D^*DK$ vertex that we use in our calculation is a $P$-wave coupling, the $D^*D_{s1}(2460)K$ vertex could receive an $O(m_D/E_K)$ enhancement. While this is not relevant for the low-$q^2$ region, such a coupling would potentially lead to sizable effects in the high-$q^2$ region. However, in the limit of soft kaon momentum the Goldstone-boson nature of the kaon implies 
 a derivative (or chiral-breaking) coupling, which in turn implies a suppression. Moreover, Eq.~\eqref{eq:dC9tot} already overestimates the contribution to $\delta C_9$ from the $D^*D$ mode by a factor of $2-3$ in the high-$q^2$ region, as can be seen in Fig.~\ref{fig:matel}, hence in this region  there is room for sizable enhancements without contradicting the upper bound in Eq.~\eqref{eq:findC9}.

To conclude, we stress that while the estimate of  $DD^*$ rescattering at high-$q^2$ from VMD and HHChPT reported in Eq~(\ref{eq:ourRes}) is based on controlled (and potentially improvable) approximations, the extrapolation at low-$q^2$ and all the two-body modes discussed in this Section is only meant to provide an 
upper bound on the 
size of effect. 
Based on the considerations discussed above, 
we conclude that unaccounted-for long-distance corrections in  $B^0\to~K^0\llpair$ should not exceed $10\%$ of the short-distance contribution induced by the $O_9$ operator within the Standard Model. 
However, we stress once more that this final estimate is not the result of a rigorous (systematically improvable) calculation.

\section{Conclusions}\label{sec:conc}

In this letter, we have presented an estimate of $B^0 \to K^0 \llpair$
long-distance contributions induced by the rescattering of a pair of charmed and charmed-strange mesons. 
We estimated these contributions using an effective description in terms of meson fields, matched to data (on $B \to D^*D_s, D^*_sD$), and theoretical constraints from HHChPT. We further included well-motivated hadronic form factors to extrapolate the result to the whole kinematical region. While the model itself is not meant to be taken too seriously, particularly away from the high-$q^2$ endpoint where it reduces to HHChPT with vector meson dominance, it is well-behaved enough and sufficiently constrained by data to provide a realistic 
estimate of this class of rescattering amplitudes. 

Our analysis partially confirms the findings of Ref.~\cite{Ciuchini:2022wbq,Ciuchini:2020gvn} that these rescattering contributions, usually neglected in theory-driven estimates of $B\to~K^{(*)}\mu^+\mu^-$
amplitudes, are relatively flat in $q^2$ (far from the 
narrow charmonium states) and can mimic a short-distance effect. On the other hand, our explicit estimate indicates that these long-distance contributions are not large: the one from the $D^*D_s$ and $D^*_sD$ intermediate states,
that we have estimated explicitly, does not exceed a few percent relative to the short-distance amplitude. 

A na\"ive estimate of other exclusive 
two-body topologies of this type indicates that the maximal correction  in the extraction of $C_9$ from data in the low-- or high--$q^2$ regions should not exceed $10\%$. However, this estimate is admittedly quite rough, being not based on the explicit estimate of rescattering channels other than the  $D^*D_s(D^*_sD)$ one. The impact of electric-dipole interactions has also been neglected.

Despite the limitations mentioned above,
the result we have presented shows that a quantitative estimation of these rescattering effects is possible, and can be improved in the future by considering all relevant channels and refining the description of the relevant hadronic form factors.
A further interesting outcome of our analysis is 
that it re-asserts the claim of Ref.~\cite{Bordone:2024hui} that comparing the extraction of $C_9$ in different $q^2$ windows provides a useful strategy to further constrain the maximal size of these long-distance contributions directly from data. 

\subsection*{Acknowledgements}

We thank Nico Gubernari, Luca Silvestrini, and Mauro Valli for their useful comments and discussions.  
This project has received funding from the European Research Council~(ERC) under the European Union's Horizon~2020 research and innovation programme under grant agreement 833280~(FLAY), and by the Swiss National Science Foundation~(SNF) under contract~200020\_204428.

\bibliography{references}
\end{document}